\documentclass[]{article}
\usepackage{amsmath}
\usepackage{cite}
\title{Note on asymptotic symmetry of massless scalar field at null infinity}
\author{Dejan Simi\' c\footnote{Email:dsimic@ipb.ac.rs}\\Institute of Physics, University of Belgrade,\\ Pregrevica 118, 11080 Belgrade-Zemun, Serbia}

\begin{document}

\maketitle

\begin{abstract}
In this article we address the question of asymptotic symmetry of massless scalar field at null infinity. We slightly generalize notion of asymptotic symmetry in order to make sense for the theory without gauge symmetry. Derivations of the results are done in two different ways, using Hamiltonian analysis and using covariant phase space. The results are in agreement with the ones previously obtained by various authors for dual 2-form field and with the results obtained starting from scalar soft theorem. 
\end{abstract}

\section{Introduction}
Asymptotic symmetries at null infinity \cite{1}have attracted substantial amount of attention in last decade, see \cite{2} for overview. Reason being their connection with soft theorems and memory effect.

A massless scalar field has a soft theorem associated to it \cite{3} and we expect that there is the corresponding asymptotic symmetry. A problem arises when we try to understand what this asymptotic symmetry is because there is no gauge symmetry in the theory.

There are many approaches to understand asymptotic symmetry of massless scalar and its connection to soft scalar theorem. Starting from the the soft scalar theorem the authors in \cite{4} derived what the asymptotic symmetry should be and generalized the result to all even dimensions \cite{5}. Scalar field is dual to 2-form field which is the theory with gauge symmetry. By passing on the dual 2-form formulation of scalar field the standard approach is used to derive the asymptotic symmetry \cite{6} and \cite{7}. This lack of symmetry in one and presence in other formulation of the theory led some to the conclusion that symmetry of theory is union of symmetries in all formulations of the theory \cite{8}. This is an unsatisfactory solution because dual theories should have same number of symmetries. 
To better understand this problem an approach that relies on compact extra dimensions \cite{d1} is proposed. 
There is also related work in spacelike infinity based on the invariance of the symplectic form under Poincare transformations \cite{10} which, also, did not yield symmetry for the scalar field but recovered results obtained at null infinity for 2-form field. This work suggests that for scalar field search for missing symmetry is not a problem of finding boundary degrees of freedom as it was in electromagnetism \cite{9}. This approach is generalized to any massless boson in \cite{d2} and \cite{d3} with the same conclusion that there is no apparent symmetry for a massless scalar.

Differences between our and previous approaches is that we work with scalar field theory and not with dual 2-form or some extended formulation and propose generalization of the notion of asymptotic symmetry at null infinity. The standard understanding of asymptotic symmetry starts with gauge symmetry, with some asymptotic conditions imposed, the part of gauge symmetry that respects the asymptotic conditions is allowed. The next step is the derivation of the associated conserved charges, that are given as integral over corner at infinity, generically some allowed gauge transformations will have identically zero charges and they are called trivial gauge transformations. Allowed gauge transformations with nonzero charges are an asymptotic symmetry of the theory or in other words allowed modulo the trivial gauge transformations.
To extend the standard notion of asymptotic symmetry we start from the observation that all the calculations are done asymptotically and that everything is ultimately about charges and their conservation. Then, it is natural to propose what the asymptotic symmetry for any theory, with or without gauge symmetry, should be. The proposed generalization is as follows. If an asymptotic transformation, not a priory defined in whole spacetime, can be represented with non zero and conserved charge then it is asymptotic symmetry of the theory. Conservation of charges in this setup can have a more subtle meaning, for example conservation for symmetries at null infinity means that charges at past and null infinity are equal and this is established by more detailed inspection of the properties of the solutions. Starting from this definition of asymptotic symmetry we can, at least in principle, obtain globally defined transformation if it is consistent with equations of motions. The globally defined transformation is the one which maps solutions onto solutions and in asymptotic region coincides with the asymptotic symmetry, this would explicitly be done by boundary-bulk propagator. Notice, that when dealing with gauge theory we can think of asymptotic symmetry in this way, with the additional need to fix trivial gauge transformations in order to obtain unique boundary-bulk propagator.
This generalization of asymptotic symmetry seems justified for null infinity, as will be demonstrated in the main part of the paper, but to see is it possible to apply it to more general case requires more investigation.

We demonstrate in the rest of the article how this extended asymptotic symmetry can be obtained using the example of massless scalar field. We do it both in Hamiltonian and covariant phase space formalism as these are two dominant approaches both with its own pros and cons.  
\section{Hamiltonian charges}
\setcounter{equation}{0}
\subsection{Coordinates}
We will work in four spacetime dimensions although it is trivial to see that results hold in other dimensions.
We use mostly minus convention for the Minkowki metric.
In Hamiltonian approach we use use light cone coordinates $u=t-r$, $v=t+r$ and $x^a$ are the coordinates on the sphere. The metric is given by
\begin{equation}
	ds^2=dudv-g_{ab}dx^adx^b\\,
\end{equation}
with
\begin{equation}
	g_{ab}=(\frac{u-v}{2})^2\gamma_{ab}\\,
\end{equation}
where $\gamma_{ab}$ is the metric on the unit sphere. Explicit form of the metric on the sphere $\gamma_{ab}$ is not important for the details of our analysis and can be taken is the standard form or in complex coordinates as in \cite{1} and \cite{2}. Future null infinity $J^+$ is reached in the limit $v\rightarrow\infty$ with other coordinates fixed. While, past null infinity $J^-$ we obtain in the limit $u\rightarrow-\infty$ with other coordinates fixed.

In the covariant phase space formalism we can use light cone coordinates or outgoing coordinates for future null infinity with the metric
\begin{equation}
	ds^2=du^2+2dudr-r^2\gamma_{ab}dx^adx^b\\,
\end{equation}
where in limit $r\rightarrow\infty$ we reach future null infinity $J^+$.
The metric in ingoing coordinates, that are suitable for past null infinity, is given by
\begin{equation}
	ds^2=dv^2-2dvdr-r^2\gamma_{ab}dx^adx^b\\,
\end{equation}
and past null infinity $J^-$ we get in $r\rightarrow\infty$ limit.
\subsection{Canonical analysis}
Action of self-interacting massless scalar in light cone coordinates is given by
\begin{equation}
	S=\int d^4x\sqrt{-g} (\partial_u\phi\partial_v\phi-\frac{1}{2}g^{ab}\partial_a\phi\partial_b\phi -V(\phi))\\,
\end{equation}
where $V(\phi)$ is any polynomial with degree higher than 3, the reason why $\phi^3$ interaction must be excluded will be described later in text. We will focus on future null infinity $J^+$ as it is trivial to see that same calculation is valid for past null infinity $J^-$, just by changing $u$ and $v$. Massless particles evolve along the $v$ direction and for $v\rightarrow\infty$ they go to $J^+$. For this reason it is valid to take $v$ as an evolution parameter in Hamiltonian dynamics.
For more details on Hamiltonian dynamics see \cite{11} and \cite{12}.

The impulse is obtained by the standard definition
\begin{equation}
	\pi=\frac{\partial L}{\partial\partial_v\phi}=\sqrt{-g}\partial_u\phi\\. 
\end{equation}
Because the right-hand side does not contain a derivative over $v$ we get the constraint (at every point)
\begin{equation}
	\psi=\pi-\sqrt{-g}\partial_u\phi\approx 0\\.
\end{equation}
The Poisson bracket of constraints at different points is
\begin{equation}
	\Omega(v,x,x')=\{\psi(v,x),\psi(v,x')\}=-2\partial_u\delta(x-x')\\,
\end{equation}
where we introduced abbreviation $x=(u,x^a)$.

The total Hamiltonian determines the evolution of the system \cite{11} and is given by
\begin{equation}
	H_T=H+\int dud^2x^a\Lambda\psi\\,
\end{equation}
where $H$ is Hamiltonian that we calculate in usual manner 
\begin{equation}
	H=\int dud^2x^a(\pi\partial_v\phi-L)=\int dud^2x^a\sqrt{-g} (\frac{1}{2}g^{ab}\partial_a\phi\partial_b\phi +V(\phi))\\.
\end{equation}
Because Hamiltonian is obtained when integrating over a Cauchy surface, and $v=const$ is a Cauchy surface only when $v$ goes to infinity, the rest of analysis is valid only in immediate vicinity of $J^+$.
The consistency condition of the constraint
\begin{equation}
	\{\psi,H_T\}=0 \\,
\end{equation}
gives the solution for multiplier $\Lambda$ up to arbitrary $u$ independent function. The presence of an arbitrary function means that there is first-class constraint hidden in the second-class constraint $\psi$. An alternative way to see that there is first-class constraint hidden among the second class is by looking for the presence of a zero mode in $\Omega$(v,x,x'), see \cite{12}. The eigenvalue equation takes form
\begin{equation}
	\int d^3x'\Omega(v,x,x')k(x')=2\partial_uk(x)\\,
\end{equation}
and $k(x)$ is a zero mode if and only if it does not depend on $u$. 

\subsection{Charges}
In order to construct the charge we will adopt the approach used in \cite{14} where the same situation appeared. We construct the generator using the whole constraint $\psi$ and not only the first-class part but we multiply it with the $u$ independent function $\Lambda$ and not completely arbitrary one
\begin{equation}
	\Psi=\int dud^2x^a \Lambda\psi\\,
\end{equation}
this is enough to select the first class from the full constraint, see \cite{14} for more details.

We demand that variation of the generator is well-defined meaning that there is no surface term
\begin{equation}
	\delta \Psi=\int A\delta\phi+B\delta\pi\\,
\end{equation}
which leads to the need to add surface term to the generator \cite{13}
\begin{equation}
	\tilde{\Psi}=\Psi+Q=\int \Lambda \pi\\,
\end{equation}
where the surface term is given by
\begin{equation}
	Q=\int dud^2x^a\sqrt{g}\Lambda\partial_u\phi\\,
\end{equation}
and represents the charge associated to the transformation.
The asymptotic behavior of scalar field near null infinity is
\begin{equation}
	\phi\propto\frac{2\varphi(u,x^a)}{v}+\mathcal{O}(\frac{1}{v^2})\\,
\end{equation}
this is the usual $\frac{\varphi}{r}$ asymptotic used in \cite{4} and \cite{10} just in light cone coordinates. We immediately derive the behavior of $\Lambda$ from
\begin{equation}
	\delta_\Lambda\phi=\{\phi,\tilde{\Psi}\}=\Lambda\\,
\end{equation}
and get
\begin{equation}
	\Lambda\propto\frac{2\lambda(x^a)}{v}+\mathcal{O}(\frac{1}{v^2})\\.
\end{equation}
Substituting the asymptotic behaviors of the field and parameter $\Lambda$ into the expression for charge, after simple algebra, we obtain 
\begin{equation}
	Q=\int dud^2x^a\sqrt{\gamma}\lambda(x^a)\partial_u\varphi(u,x^a)=-\int d^2x^a\sqrt{\gamma}\lambda(x^a)\varphi(-\infty,x^a)\\.
\end{equation}
The last equality follows from the fact that there are no massive particles because then the field $\varphi$ goes to zero at $u\rightarrow\infty$, see \cite{2} and \cite{4}.
The result is in agreement with the expressions for charge obtained in \cite{4,5,6,7}.

We can repeat the same calculation at past null infinity $J^-$ and obtain the charge
\begin{equation}
	Q_{J^-}=\int dvd^2x^a\sqrt{\gamma}\lambda(x^a)\partial_v\varphi(v,x^a)=\int d^2x^a\sqrt{\gamma}\lambda(x^a)\varphi(\infty,x^a)\\.
\end{equation}
At first glance this two charges are unrelated, as the parameters $\lambda$ at past and future null infinity are not connected, and it's not obvious how to establish conservation in any sense. Careful inspection of the equations of motion \cite{4} reveals that the asymptotic values of field at past and future null infinity are connected. Namely, they are equal for antipodal points that approach spatial infinity
\begin{equation}
	\varphi_{J^+}(u=-\infty,x^a)=\varphi_{J^-}(v=+\infty,-x^a)\\.
\end{equation}
The reason for this can be traced back to the discontinuity of boosts at spatial infinity \cite{2}. If we impose the antipodal matching condition $\lambda_{J^+}(x^a)=-\lambda_{J^-}(-x^a)$ on parameter $\lambda$, then we obtain equality of charges at past and future null infinity and retrieve the conservation of charges.

\section{Covariant phase space}
\setcounter{equation}{0}
Now we analyze the symmetry at null infinity of massless scalar field in the covariant phase space approach, for an introduction see \cite{15}. How to systematically construct charges is well elaborated on in \cite{16}.
\subsection{Symplectic form}
The starting point is variation of the action
\begin{equation}
	S=\int \sqrt{-g}d^4x (\frac{1}{2}g^{\mu\nu}\partial_\mu\phi\partial_\nu\phi -V(\phi))\\,
\end{equation}
which is easy to calculate
\begin{equation}
	\delta S=\int d^4x(-\partial_\mu(\sqrt{-g}g^{\mu\nu}\partial_\nu\phi)\delta\phi-\sqrt{-g}\frac{\partial V}{\partial\phi}\delta\phi+\partial_\mu(\sqrt{-g}g^{\mu\nu}\partial_\nu\phi\delta\phi))\\,
\end{equation}
from which we obtain the equation of motion (EOM)
\begin{equation}
	\partial_\mu(\sqrt{-g}g^{\mu\nu}\partial_\nu\phi)+\sqrt{-g}\frac{\partial V}{\partial\phi}=0\\,
\end{equation}
as well as the presymplectic potential
\begin{equation}
	\theta^\mu= \sqrt{-g}g^{\mu\nu}\partial_\nu\phi\delta\phi\\.
\end{equation}
The symplectic form on spacelike surfaces is given by the standard formula
\begin{equation}
	\Omega=\delta\int_\Sigma n_\mu\theta^\mu\\,
\end{equation}
where $\Sigma$ is Cauchy surface with the orientation given by unit normal $n_\mu$ that points toward the future, see \cite{16}. Generally, the symplectic form is surface term obtained by applying Stokes theorem on $\int d^4x \partial_\mu\delta\theta^\mu$ and keeping only the relevant surface term. For future null infinity only terms at $J^+$ and at past null infinity only terms at $J^-$ with the additional change of sign because normal points to the "past".

At future null infinity, assuming the same asymptotic behavior of the scalar field as before, the symplectic form is given by
\begin{equation}
	\Omega=\int_{J^+}dud^2x^a\sqrt{\gamma}\delta\partial_u\varphi\delta\varphi\\.
\end{equation}
The same result, only with $u$ replaced by $v$, holds at past null infinity.
\subsection{Symmetry at null infinity}
The symplectic form is always invariant under the transformation
 \begin{equation}
 	\delta_\lambda\varphi=\lambda
 \end{equation}
where $\lambda$ is field independent $\delta\lambda=0$. This is the trivial invariant of the symplectic form that is always present and does not automatically imply symmetry. An additional condition that transformation must fulfill to be a symmetry is that it maps solution onto solution.

The equation of motion in $(u,r,x^a)$ coordinates is
\begin{equation}
\sqrt{\gamma}r^2\partial_u\partial_r\phi+\sqrt{\gamma}\partial_r(r^2\partial_u\phi)+\partial_a(\sqrt{\gamma}\gamma^{ab}\partial_b\phi)+\sqrt{\gamma}r^2\frac{\partial V}{\partial\phi}=0\\.
\end{equation}
Assuming the same asymptotic behavior of the scalar field $\frac{\varphi}{r}$ as before, we see that if there was $\frac{g}{3}\phi^3$ term in the potential $V$, the leading term is of order $\mathcal{O}(1)$ and reads
\begin{equation}
	g\varphi^2=0\\,
\end{equation}
namely $\phi^3$ interaction makes asymptotic theory trivial. Next orders of expansion of the EOM give the solution for subleading terms as functions of free data $\varphi$. Completely analogous equations hold in advanced coordinates $(v,r,x^a)$ at past null infinity.
 
The invariance of symplectic form under transformation is equivalent to the claim that we can obtain the charge associated to that transformation via the following equation \cite{16}
\begin{equation}
		\delta Q=-I_{X_\lambda}\Omega\\,
\end{equation}
where $I_{X_\lambda}$ is contraction that acts as $I_{X_\lambda}\delta\varphi=\lambda$.
Charge $Q$ is easily calculated using previous equation
\begin{equation}
	Q=\int dud^2x^a\sqrt{\gamma}(\lambda\partial_u\varphi-\partial_u\lambda\varphi)\\,
\end{equation}
where there is an additional term proportional to $\partial_u\lambda$ in comparison to the result for charge in the Hamiltonian approach. We point out again that charge is not useful if there is not some kind of conservation associated to it. We see that if parameter $\lambda$ is $u$ independent we get the same result for charge as in the Hamiltonian approach and it is conserved with the addition of the antipodal matching condition. We argue that for any $\lambda$ that does depend on $u$ this is not possible. By doing partial integration the charge can be transformed into
\begin{equation}
	Q_{J^+}=-\int d^2x^a\sqrt{\gamma}\lambda_{J^+}\varphi_{J^+}(u=-\infty)-2\int dud^2x^a\sqrt{\gamma}\partial_u\lambda_{J^+}\varphi_{J^+}\\,
\end{equation}
and analogous for past null infinity $J^-$. The first term is the same as Hamiltonian charge that is conserved due to the antipodal matching condition. This implies that second term must be conserved separately as
\begin{equation}
	\int dud^2x^a\sqrt{\gamma}\partial_u\lambda_{J^+}\varphi_{J^+}=\int dvd^2x^a\sqrt{\gamma}\partial_v\lambda_{J^-}\varphi_{J^-}\\.
\end{equation}
We expect that fields at past and future null infinity can be related via transformation that should be nonlinear due to the interaction
\begin{equation}
	\varphi_{J^+}=\int dvd^2x'^aS(u,x,v,x';\varphi_{J^-})\varphi_{J^-}\\,
\end{equation}
which after substitution into the above equation leads to
\begin{equation}
	\int dud^2x^aS(u,x,v,x';\varphi_{J^-})\partial_u\lambda_{J^+}=\partial_v\lambda_{J^-}(\varphi_{J^-})\\.
\end{equation}
This means that starting from the field-independent parameter $\lambda_{J^+}$ we get field dependent parameter $\lambda_{J^-}$ at past null infinity. This contradicts the starting, and crucial, assumption for construction of the charge, that parameter $\lambda$ is field-independent. Consequently, we are forced into taking $u$ independent $\lambda_{J^+}$ and $v$ independent $\lambda_{J^-}$, in agreement with the Hamiltonian approach.
\section{Discussion}
We derived asymptotic symmetry of massless scalar field at null infinity directly and not by passing to dual 2-form field formulation.  

The first derivation is Hamiltonian and relies crucially on the presence of constraints in the theory. In scalar field case constraints appear only is special coordinates, we worked in light cone coordinates, and in most of other coordinates the symmetry is completely hidden. 

Covariant phase space offers more direct and simple way of deriving symmetry. We search for transformations that leave the symplectic form at null infinity invariant,  then we the construct charge via variational equation and if and only if it is conserved ,meaning that it is the same at past and null infinity, the transformation is really asymptotic symmetry. This gives a computational approach that can be applied to any theory and unravel the hidden asymptotic symmetries. This is the topic for further research.

Besides the direct derivation of asymptotic symmetry another unanswered question is how do these charges act. We offer our view on this in the context of the extended notion of asymptotic symmetry.

Globally defined symmetry maps solutions onto solutions. Starting from this obvious claim we can demand that action of asymptotic symmetry is extended on to the whole spacetime in a way that satisfies this requirement. This can be done when asymptotic symmetry shifts initial and final conditions at null infinity in a way that is consistent with EOM. For an explicit expression of the action of symmetry transformation we would need boundary-bulk propagator i.e. explicit solution with given initial or final conditions at null infinity. Because in the case of the scalar field antipodal matching conditions for field $\varphi$ and parameter $\lambda$ have opposite signs asymptotic symmetry cannot be extended into the bulk and is defined only asymptotically. 

An open problem that remains is the derivation of the symmetry at spatial infinity.  The approach of \cite{10} shows that there is no justification for adding boundary degrees of freedom at spatial infinity for the scalar field, that is necessary ingredient in their approach and it seems like the same holds for the approach of this article. 
We must conclude that there is still a long way to go if we want to fully understand symmetries in field theory. 
\section*{ACKNOWLDGMENTS}
This work was supported by the Ministry of Education, Science and Technological Development of the Republic of Serbia.

\end{document}